\documentclass[aps,prl,preprint,superscriptaddress]{revtex4-1}

\usepackage{graphicx}
\usepackage{natbib}
\usepackage[breaklinks]{hyperref}
\usepackage{amsmath}
\usepackage{color}
\usepackage{gensymb}

\begin{document}

\title{Varying temperature and silicon content in nanodiamond growth: effects on silicon-vacancy centers}
\author{Sumin Choi}
\affiliation{Data Storage Institute, Agency for Science, Technology and Research, 138634 Singapore}
\author{Victor Leong}
\affiliation{Data Storage Institute, Agency for Science, Technology and Research, 138634 Singapore}
\author{Valery A. Davydov}
\affiliation{L.F. Vereshchagin Institute for High Pressure Physics, the Russian Academy of Sciences, Troitsk, Moscow, 142190 Russia}
\author{Viatcheslav N. Agafonov}
\affiliation{GREMAN, UMR CNRS-7347, Universit\'e F. Rabelais, 37200 Tours, France}
\author{Marcus~W.O.~Cheong}
\affiliation{Data Storage Institute, Agency for Science, Technology and Research, 138634 Singapore}
\author{Dmitry A. Kalashnikov}
\affiliation{Data Storage Institute, Agency for Science, Technology and Research, 138634 Singapore}
\author{Leonid A. Krivitsky}
\email[]{Leonid-K@dsi.a-star.edu.sg}
\affiliation{Data Storage Institute, Agency for Science, Technology and Research, 138634 Singapore}
\date{\today}

\begin{abstract}
  Nanodiamonds containing color centers open up many applications in quantum information processing, metrology, and quantum sensing.
  In particular, silicon vacancy (SiV) centers are prominent candidates as quantum emitters due to their beneficial optical qualities.
  Here we characterize 
  nanodiamonds produced by a high-pressure high-temperature method without catalyst metals,
  focusing on two samples with clear SiV signatures.
  Different growth temperatures and relative content of silicon in the initial compound
  between the samples altered their nanodiamond size distributions and abundance of SiV centers.  
  Our results show that nanodiamond growth can be controlled and optimized for different applications.

\end{abstract}

\maketitle
\section{Introduction}

Color centers in diamond have emerged as important quantum emitters for a broad range of applications including bioimaging~\cite{simpson2014vivo,sotoma2016selective,wu2017nanodiamonds}, sensing~\cite{cooper2014time,tzeng2015time}, and quantum nanophotonics~\cite{sipahigil:science2016,schroder2016quantum}.
One important example is the silicon vacancy (SiV) center, which has been an active focus of research in recent years
due to its attractive optical properties~\cite{neu2012photophysics,neu2013low,jantzen2016nanodiamonds,li2016nonblinking},
including 
high brightness, narrow homogenous distribution, stable single photon emission with near-transform-limited linewidths, and minimal spectral diffusion.
Its zero-phonon line (ZPL) at 737\,nm contains $\sim$70\% of the emitted fluorescence, and inversion symmetry grants an insusceptibility to electric field fluctuations.
Recent works have also explored the applications of SiV centers based on diamond nanostructures~\cite{felgen2016incorporation,sipahigil:science2016}.

Nanodiamonds (NDs) containing color centers
can be spatially manipulated and precisely positioned for enhanced coupling to other nanophotonics structures~\cite{wolters2010enhancement,van2011deterministic} or to fibers~\cite{liebermeister2014tapered,benedikter2017cavity}.
The small size of NDs is also advantageous in bioimaging and sensing applications~\cite{perevedentseva2013biomedical,knowles2013observing},
and may enhance coherence times in the SiV centers~\cite{rogers2014all,jahnke2015electron}.
In principle, the ND composition can be optimized for different applications.
For instance, fluorescent imaging probes require a high density of emitters for increased brightness and must be stable against photobleaching,
while many quantum networking tasks require single emitters as true single-photon sources. 

Here, we explore the ability to control the ND size distribution and abundance of SiV centers by adjusting the growth temperature and relative silicon content in the initial ND growth compound.
We perform room-temperature characterization of several ND samples
and compare their physical and optical properties.

\section{Nanodiamond preparation}

The NDs with SiV centers used in this work were synthesized using a high-pressure high-temperature (HPHT) process without metal catalysts, 
based on mixtures of naphthalene~(C$_{10}$H$_8$) and tetrakis(trimethylsilyl)silane (C$_{12}$H$_{36}$Si$_5$)
with different silicon-to-carbon (Si/C) ratios in the initial compound. 
In this work, we focus on two samples, namely 
sample~A (Si/C ratio: 0.008) and sample~B (Si/C ratio: 0.05);
the remaining samples did not show clean SiV spectral signatures, and are discussed in Appendix A.
HPHT treatment of the initial homogeneous mixtures was carried out in a high-pressure apparatus of ``Toroid'' type~\cite{davydov2014production}.
The experimental procedure consists of loading the high-pressure apparatus to 8.0\,GPa, 
heating the samples up to 1300\degree C and 1450\degree C for samples~A and B, respectively,  
and short (5\,s) isothermal exposures at these temperatures. 

The obtained diamond products in both samples consist of nano- and submicron-sized diamond fractions, but with different particle size distributions.
As we are primarily interested in small NDs, we investigate only the smallest size fraction from each sample, which consists of NDs
10--30\,nm in size for sample~A, and 50--100\,nm for sample~B.
The difference in ND size distributions
can be attributed to the higher growth temperature inducing a more active cumulative recrystallization process for sample~B, which leads to larger NDs.
This has also been observed in carbon nanosystems where a hydrocarbon component is introduced~\cite{davydov2016comparative}.

In contrast to NDs grown via chemical vapor deposition (CVD) on a silicon or metal substrate~\cite{neu2013low,li2016nonblinking}, 
these samples are produced in a powder form, making them convenient for further processing and subsequent spatial manipulation, 
which is crucial for coupling to photonic nanostructures.

After extraction from the high-pressure apparatus,
both samples undergo ultrasonication and centrifugation to reduce clustering and to isolate the smallest NDs, respectively,
before being spin-coated onto a silicon substrate for further characterization.
The detailed procedure is described in Appendix A.

\begin{figure}[tbp]
\centering
  \includegraphics[width=0.6\columnwidth]{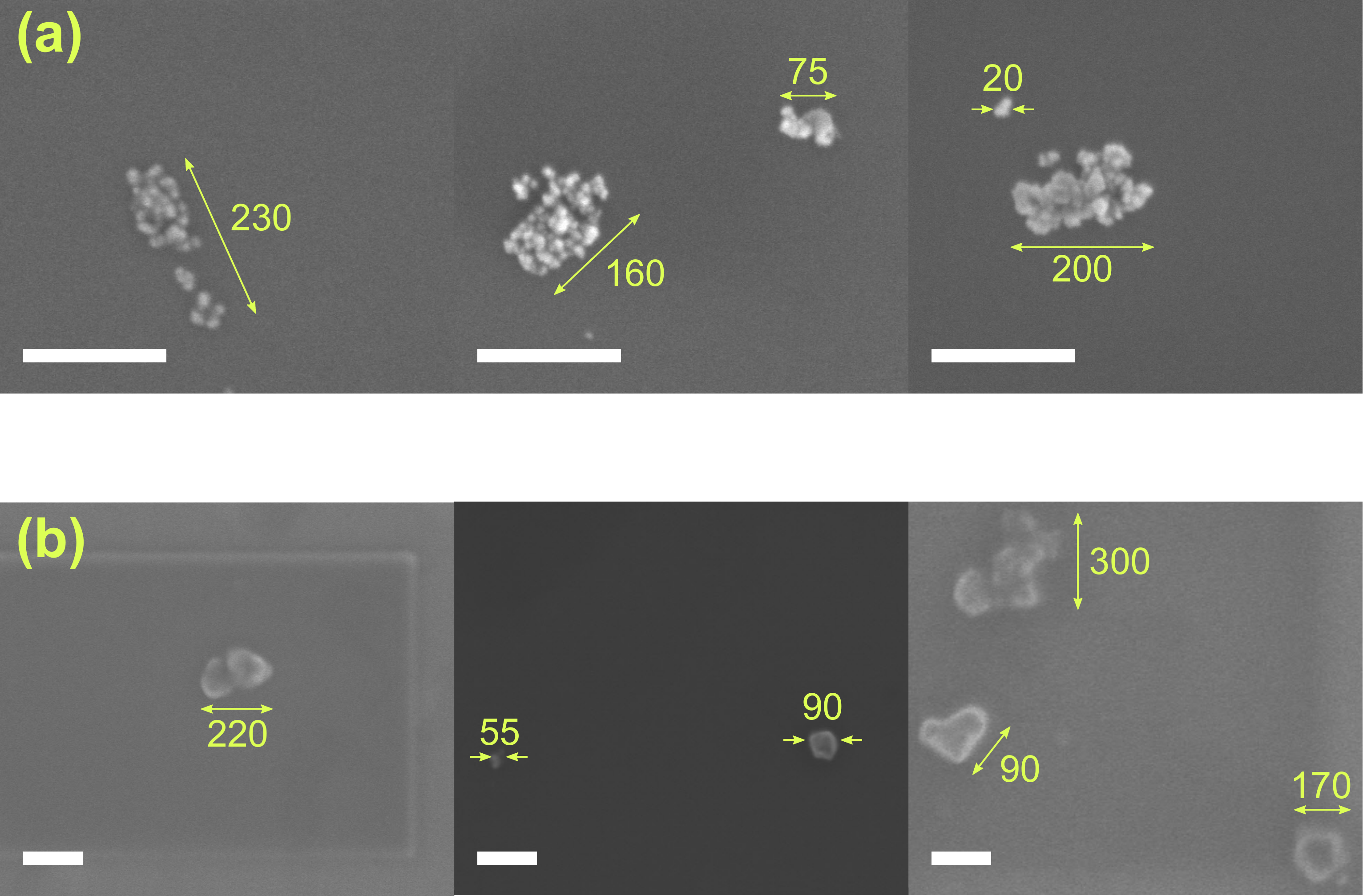}
  \includegraphics[width=0.39\columnwidth]{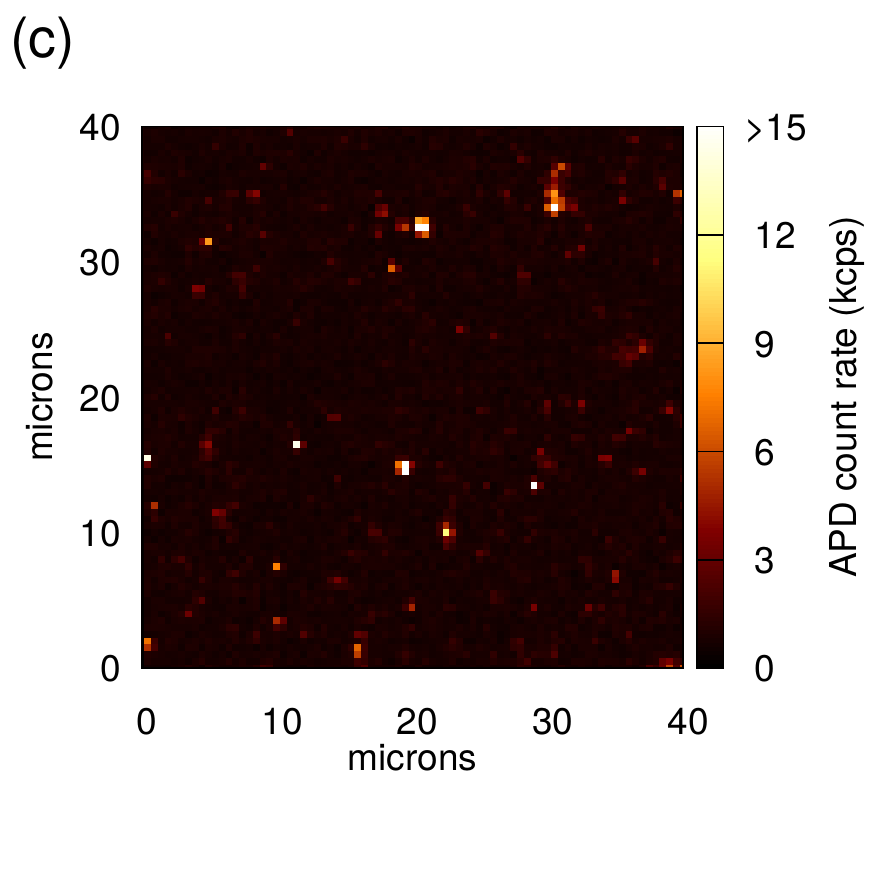}
  \caption{\label{fig:sem}
  (a),(b) SEM pictures of samples~A and B, respectively. Numbers show approximate sizes (in nm) of individual nanodiamonds or clusters. White lines are scale bars of 200\,nm length.
  (c) Typical confocal scan map with 0.5\,$\mu$m step size. Isolated bright spots indicate fluorescent emitters.
  }
\end{figure}

Despite the ultrasonication, we are unable to eliminate clustering completely; a similar issue was reported in ref.~\cite{jantzen2016nanodiamonds} with HPHT NDs. 
Some SEM images of the samples are shown in Fig.~\ref{fig:sem}a and~\ref{fig:sem}b,
revealing individual NDs and isolated clusters of up to $\sim$300\,nm in size for both samples.

\section{Experimental Setup}

The optical properties of the NDs are studied with a home-built confocal microscope at room temperature.
A detailed description of the setup can be found in Appendix B.
Briefly, off-resonant excitation light from a continuous-wave 532\,nm laser is focused onto the sample through an air objective (NA=0.95).
The emission passes through a notch filter to reject the excitation light, is collected into a single-mode fiber,
and directed to either a grating spectrometer 
for photoluminescence (PL) measurements,
or to a Hanbury-Brown and Twiss (HBT) interferometer for measurements of the $g^{(2)}$ second-order correlation function using single photon counters. 
Except for spectrometer measurements, the emission is additionally filtered by a 740\,nm narrowband filter (Semrock, bandwith 13\,nm).

\section{Results and discussion}

A typical confocal scan map of a sample is shown in Fig.~\ref{fig:sem}c. 
Isolated bright spots reveal possible SiV candidates that would be confirmed by PL measurements.
We note that it in our setup, it is not possible to resolve multiple emitters in a ND crystal or cluster within the diffraction-limited confocal detection spot.
We then measure the $g^{(2)}$ function of SiV centers to determine if they are single emitters,
and additionally analyze the polarization and saturation behavior of single emitters.

\textbf{Photostability}

Many emitters from sample~A are not stable, and 33 out of 53 emitters containing SiV centers suffered from either photobleaching or blinking under continuous excitation over a few minutes, even under low excitation power below 300\,$\mu$W.
Blinking refers to intermittent fluorescence alternating between on/off states, while photobleaching refers to a gradual but permanent loss of fluorescence that does not recover even after long waiting times.
In contrast, all 40 investigated emitters from sample~B are stable for $>$30\,mins, even under higher excitation powers of above 4\,mW.

The lack of photostability has been studied for various types of quantum emitters~\cite{choi2014single,Tran2016}.
For color centers in NDs, 
possible mechanisms inlcude charge state switching (photoionization) and the capture of electrons in surface traps~\cite{neu2012photophysics,bradac2010observation,tisler2009fluorescence}.
These effects may be more pronounced in smaller NDs due to a lack of excess electrons, 
and could explain the increased stability of sample~B over sample~A.
Further surface treatment may lead to improved photostability~\cite{bradac2010observation,cui2013increased}.

\textbf{Photoluminiscence spectra}

\begin{figure}[tbp]
\centering
  \includegraphics[width=0.6\columnwidth]{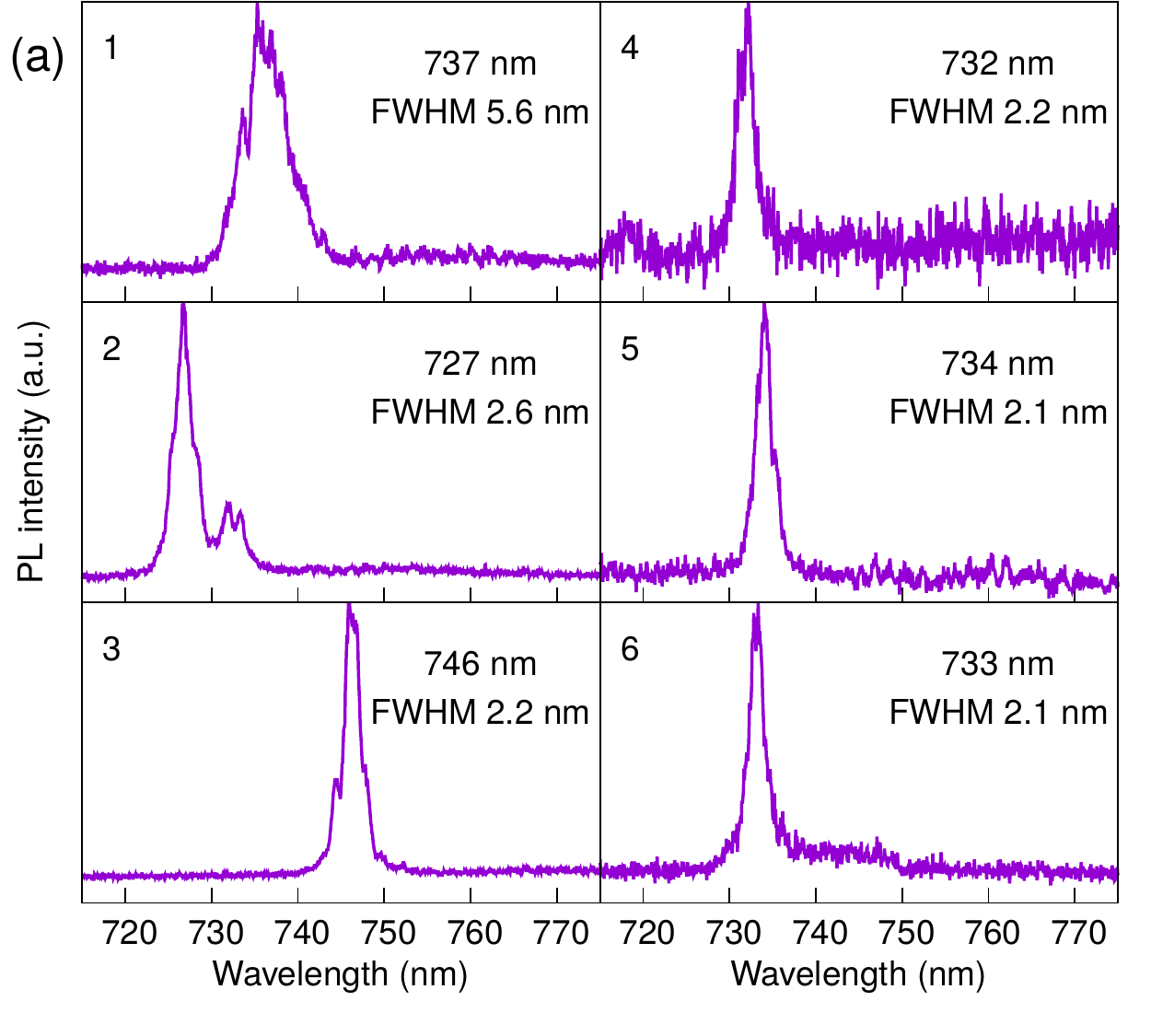}
  \includegraphics[width=0.36\columnwidth]{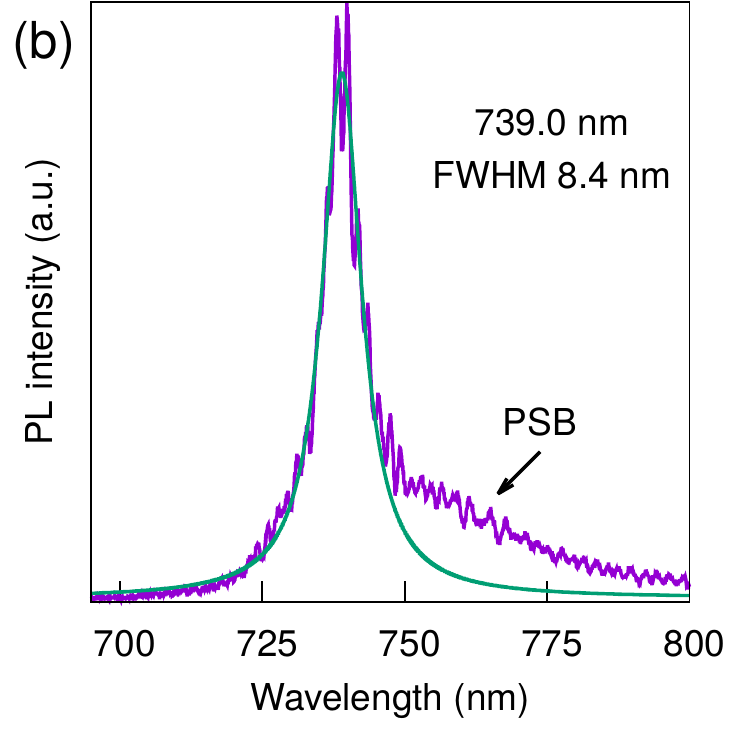}
  \caption{\label{fig:PL}
  (a) Photoluminiscence (PL) spectra of single SiV centers in sample~A, labelled as emitters 1--6. The central wavelengths of the zero-phonon line (ZPL) peak and the full-width at half-maximum (FWHM) are derived from a Lorentzian fit.
  (b) Averaged PL spectrum of 12 fluorescent spots in sample~B, overlaid with the Lorentzian fit used to measure the overall linewidth. The phonon sideband (PSB) above 750\,nm is clearly visible. }
\end{figure}

PL measurements of stable emitters in sample~A revealed fluorescence peaks scattered around the nominal ZPL wavelength of 737\,nm. 
Similar observations for SiV centers in NDs were reported elsewhere~\cite{benedikter2017cavity,neu2013low,neu2011single}, and were attributed to local strain effects in smaller NDs.
Here, we identify fluorescence peaks within the range of $737\pm10$\,nm as SiV centers; a few other peaks were observed at $>$750\,nm, but these were rejected.
Most of the peaks do not have a distinct phonon sideband (PSB), and
from Lorentzian fits we obtain full-width at half-maximum (FWHM) values of
2.1---2.9\,nm, except for two emitters with a FWHM of 3.9\,nm and 5.6\,nm. 
The spectra of six emitters labelled 1--6, later confirmed to be single SiV centers, are shown in Fig.~\ref{fig:PL}a. 

In contrast,
the emitters in sample~B showed almost identical PL spectra, with a clearly visible PSB.
Fig.~\ref{fig:PL}b shows an averaged spectrum of 12 fluorescent spots (all but one contain multiple SiV centers), with a central ZPL peak of 739.0\,nm and a FWHM of 8.4\,nm.
The sole single emitter showed a FWHM of 7.8\,nm.
Here, we cannot distinguish if the large linewidth is caused by a broad inhomogenous distribution of ZPL peaks, or if individual SiV centers have a broad ZPL spectrum.

\textbf{$g^{(2)}$ function}

To identify if a fluorescent spot consists of a single SiV center instead of multiple emitters, we consider the $g^{(2)}$ function between the two output detectors of the HBT interferometer.
We do not perform any background corrections, and fit the data to a realistic model as follows~\cite{neu2012photophysics}: 
The $g^{(2)}$ function of an ideal three-level system is given by
\begin{equation}
  g^{(2)}(\tau) = 1 - (1+\alpha) \exp (-|\tau|/\tau_1) + \alpha \exp(-|\tau|/\tau_2)\quad,
\end{equation}
where $\tau_1$ and $\tau_2$ are the lifetimes of the excited and metastable shelving states, respectively, and $\alpha$ describes the degree of bunching.
The effect of background noise can be described by $\rho^2=\frac{S^2}{(S+B)^2}$, where $S$ and $B$ are signal and background intensities, respectively, yielding
\begin{equation}\label{eq:g2m}
  g_\textrm{noisy}^{(2)}(\tau) = 1+\rho^2(g^{(2)}(\tau) - 1)\quad.
\end{equation}
Equation~(\ref{eq:g2m}) is then convolved with the independently measured timing response function of the setup, which is well-approximated by a Gaussian with $\sigma\sim0.5$\,ns (see Appendix C for details).
The final expression is used for fitting to the measured data.

\begin{figure}[tbp]
\centering
  \includegraphics[width=0.6\columnwidth]{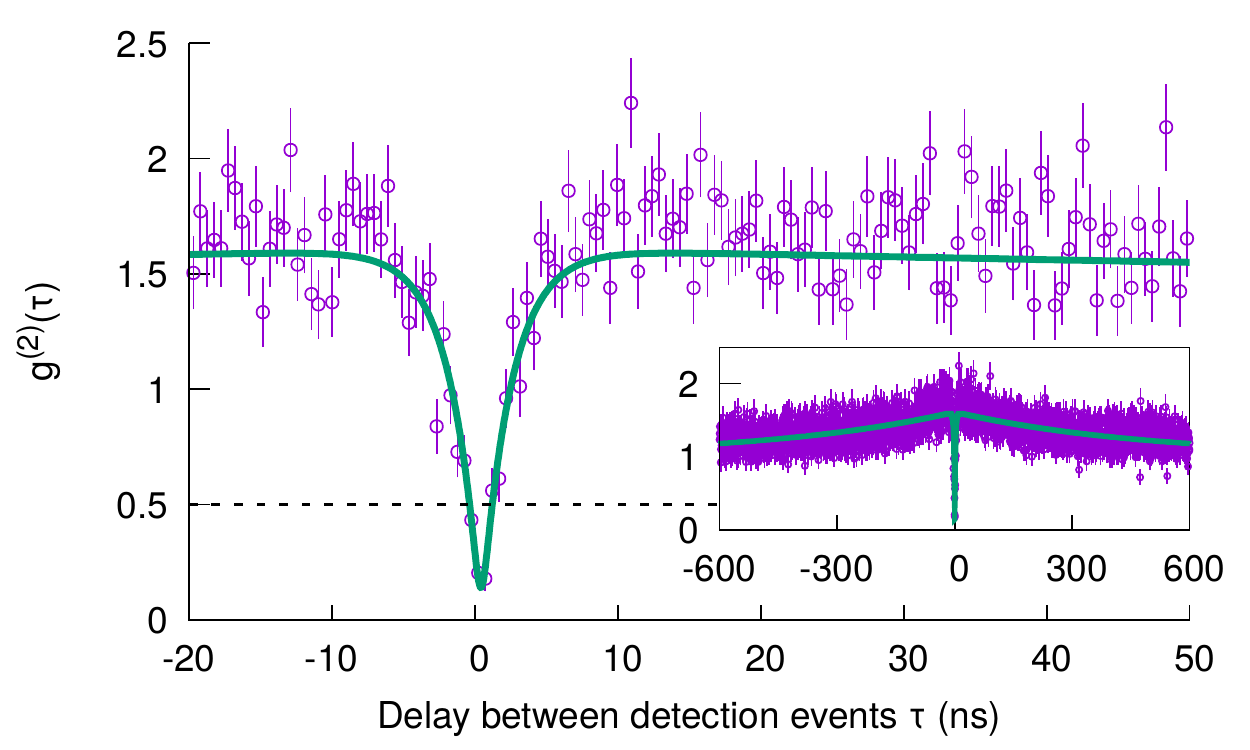}
  \caption{\label{fig:g2}
  Measured $g^{(2)}$ function of fluorescence from an emitter, with an antibunching dip of $g^{(2)}(0)<0.5$ confirming that it is a single SiV center.
  The solid line is a fit to a three-level model that accounts for detector timing jitter and background noise.
  Inset plots the same measurement at a longer time scale to show the bunching behavior. 
  }
\end{figure}

Only emitters with the characteristic antibunching signal of $g^{(2)}(0)<0.5$ can be clearly identified as single SiV centers (see Fig~\ref{fig:g2}).
For sample~A, 6 out of 20 stable emitters show $g^{(2)}(0)<0.5$ (30\%), while we observe only 1 single emitter out of 40 candidates (2.5\%) for sample~B.
We note that we are unable to quantify the proportion of NDs that do not contain any SiV centers, since they do not give a clear fluorescence signal.

Although the sizes of individual NDs in sample~A are smaller than in sample~B, we observe isolated clusters of similar sizes in both samples (Fig.~\ref{fig:sem}a,b).
As such, for each investigated emitter, we regard the volume of NDs within the confocal microscope detection spot as approximately equal for both samples,
and attribute the difference in SiV center abundance of both samples to the relative silicon content (i.e. Si/C ratio) in the initial growth compound.
We conclude that
the lower Si/C ratio for sample~A has increased the proportion of SiV-containing NDs that hosts only a single emitter.
Thus, tuning the growth conditions of NDs can aid the production of NDs with an optimized abundance of SiV centers, depending on the intended application.

From the $g^{(2)}$ fits, we are also able to extract the lifetime of the radiative transition $\tau_1$, which ranges from $0.9\pm0.2$\,ns to $3.8\pm0.2$\,ns for all the investigated single emitters.
The range of values is comparable to other reported values of $\tau_1$ for SiV centers in NDs~\cite{jantzen2016nanodiamonds,li2016nonblinking}.

\textbf{Fluorescence polarization of single emitters}

\begin{figure}[tbp]
\centering
  \includegraphics[width=0.9\columnwidth]{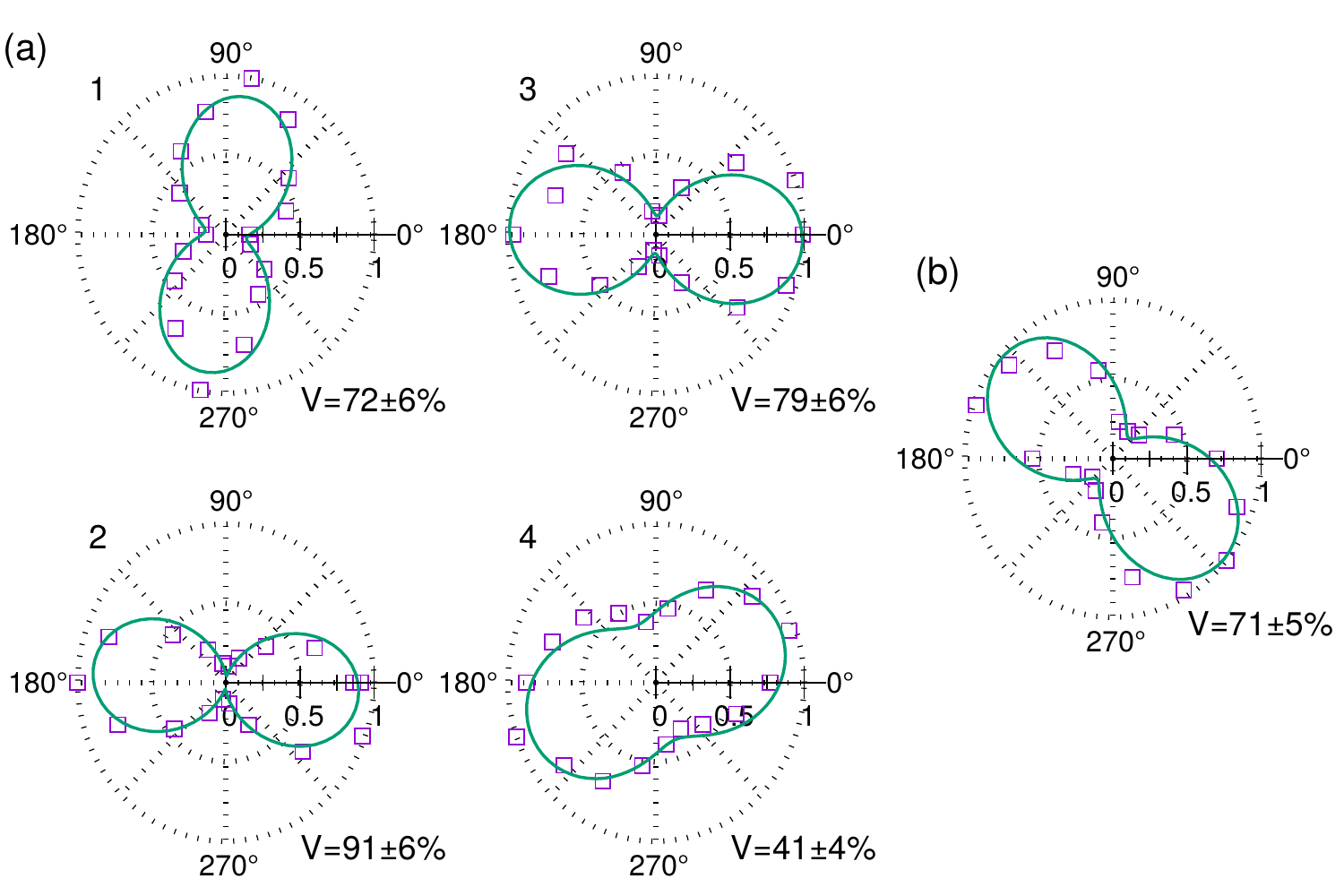}
  \caption{\label{fig:pol}
  Normalised polar plots of fluorescence count rate $I$ of (a) single emitters 1--4 from sample~A and (b)~the single emitter from sample~B, as a function of the rotation angle $\theta$ of a polarizer placed after the dichroic beamsplitter. Solid lines show fits to a cos$^2\theta$ model, from which we obtain the visib1ility $V=(I_\textrm{max}-I_\textrm{min})/(I_\textrm{max}+I_\textrm{min})$.
  }
\end{figure}

We then analyze the polarization of the emitted fluorescence of the single SiV centers by placing a polarizer after the dichroic beamsplitter, and measuring the fluorescence count rate $I$ as a function of the rotation angle of the polarizer. 
The emission of SiV centers is known to be linearly polarized~\cite{neu2011single}, and the polarization contrast can be described by the visibility
\begin{equation}
  V=\frac{I_\textrm{max}-I_\textrm{min}}{I_\textrm{max}+I_\textrm{min}}\quad.
\end{equation}
Unfortunately, emitters 5 and 6 from sample~A were bleached during this measurement. 
The results for the other single emitters are shown in Fig.~\ref{fig:pol}.

Except for one of the single emitters from sample~A, the visibility $V$ is fairly high at $>$70\%. 
The lack of full visibility can be attributed to several reasons.
The dichroic beamsplitter induces polarization changes in the transmitted light, and although we correct for the polarization-dependent transmission, the dichroic beamsplitter causes an additional loss of linear polarization of $\sim$10\%~\cite{neu2011single}.
Besides, polarization anisotrpy due to imaging from a high NA objective~\cite{fourkas2001rapid,neu2011single}, background luminescence from the diamond material, and contributions from another distant, weakly excited emitter can also degrade polarization contrast.
As such, in applications where high visibility is critical, a polarizer can be used to project the fluorescence onto an optimal linear polarization.

\textbf{Saturation behavior of single emitters}

\begin{figure}[tbp]
\centering
  \includegraphics[width=0.85\columnwidth]{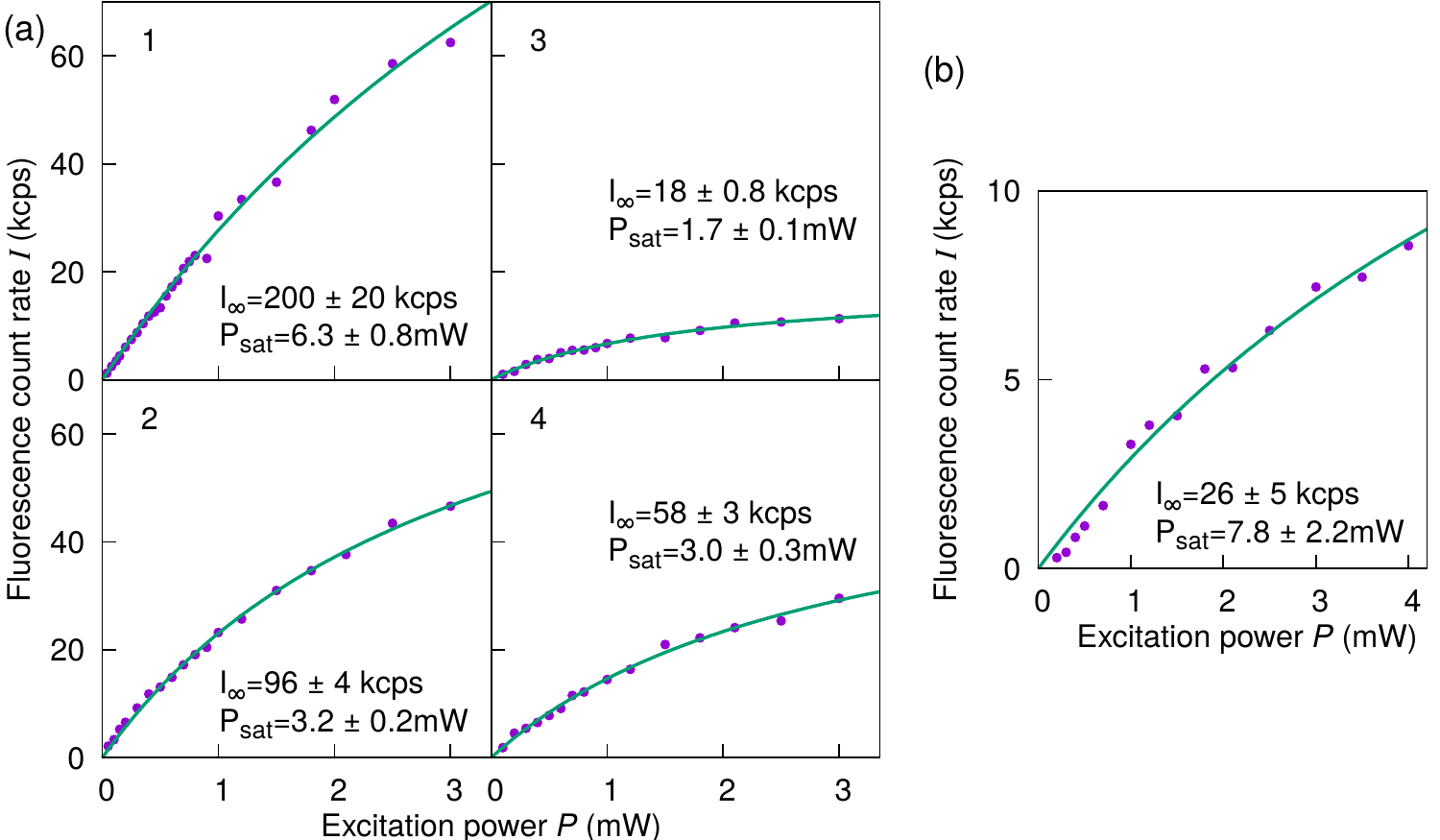}
  \caption{\label{fig:sat}
  Fluorescence count rate $I$ of (a) single emitters 1--4 from sample~A and (b)~the single emitter from sample~B, as a function of the power $P$ of the 532\,nm excitation light. 
  The data is corrected for background counts.
  Solid lines represent fits to the function $I=I_\infty\frac{P}{P+P_\textrm{sat}}$, from which we extract the maximum count rate $I_\infty$ and saturation power $P_\textrm{sat}$.
  The poor fit leading to a large uncertainty in $I_\infty$ and $P_\textrm{sat}$ in (b) might be due to an onset of photobleaching at higher excitation powers.
  }
\end{figure}

The saturation behavior of the single emitters can be described by the equation
\begin{equation}\label{eq:sat}
  I=I_\infty\frac{P}{P+P_\textrm{sat}}\quad,
\end{equation}
where $I_\infty$ is the maximum fluorescence count rate, $P$ is the excitation power, and $P_\textrm{sat}$ is the saturation power.
In our measurements, we first maximize $I$ by
rotating a half-wave plate in the excitation beam path,
then recording $I$ as a function of the excitation beam power (see Fig.~\ref{fig:sat}).
The data is corrected for the background count rate measured at a nearby spot on the substrate without any fluorescent NDs, then fitted to equation~\ref{eq:sat}.

The observed saturation behavior varied greatly between the single emitters, with the fitted $I_\infty$ values ranging from 18\,kcps to 200\,kcps, and $P_\textrm{sat}$ ranging from $1.7\pm0.1$\,mW to $7.8\pm2.2$\,mW. 
We  
note that we were not able to fully observe the fluorescence above $P_\textrm{sat}$ due to the onset of photobleaching at higher excitation powers, which might have caused the poor fit leading to a large uncertainty in $I_\infty$ and $P_\textrm{sat}$ for the single emitter from sample~B (see Fig.~\ref{fig:sat}b).
A choice of a longer excitation wavelength or resonant excitation could have allowed for more efficient excitation and a lower $P_\textrm{sat}$~\cite{neu2011single,li2016nonblinking}.

\begin{table}[htbp]
\setlength{\tabcolsep}{10pt}
\caption{\label{tab:table1}
  Summary of the observed single SiV centers. We were not able to measure the polarization and saturation behavior of emitters 5 and 6 from sample 1.
  }

  \begin{tabular}{cccccccc}
  Sample / emitter & ZPL / FWHM (nm) & $\tau_1$ (ns) & $V$ (\%) & $I_\infty$ (kcps) & $P_\textrm{sat}$ (mW) \\
  \hline
  A / 1 & 737 / 5.6 & $1.7\pm0.2$ & $72\pm6$ & $200\pm20$ & $6.3\pm0.8$ \\
  A / 2 & 727 / 2.6 & $1.1\pm0.2$ & $91\pm6$ & $96\pm4$ & $3.2\pm0.2$ \\
  A / 3 & 746 / 2.2 & $2.5\pm0.2$ & $79\pm6$ & $18\pm0.8$ & $1.7\pm0.1$ \\
  A / 4 & 732 / 2.2 & $3.8\pm0.2$ & $41\pm4$ & $58\pm3$ & $3.0\pm0.3$ \\
  A / 5 & 734 / 2.1 & $3.7\pm0.4$ & - & - & - \\
  A / 6 & 733 / 2.1 & $ 3.1\pm0.8$ & - & - & - \\
  B / 1 & 739 / 7.8 & $0.9\pm0.2$ & $71\pm5$ & $26\pm5$ & $7.8\pm2.2$
\end{tabular}
\end{table}

\section{Conclusion}
In conclusion, we have characterized SiV centers in HPHT ND samples.
The lower growth temperature for sample~A has led to smaller ND sizes compared to sample~B.
For sample~A which is obtained from a mixture with low silicon content (Si/C ratio: 0.008), 
among NDs that show a SiV spectral signature, we observe a 30\% fraction (6 out of 20 candidates) of NDs that contain a single emitter.
This is roughly ten times higher than in sample~B which is obtained from a mixture with high silicon content (Si/C ratio: 0.05), 
where the corresponding fraction is 2.5\% (1 out of 40 candidates).
We summarize the observations of all the single emitters in Table~\ref{tab:table1}.
Our results demonstrate that varying the synthesis parameter and the doping impurity content in the initial growth compound
can effectively influence the ND size distribution and the abundance of single photon emitters.
This opens up possibilities for targeted synthesis of diamond materials for different applications.

\bibliographystyle{apsrev4-1}
\bibliography{myref.bib}

\newpage

\section{Appendix A: Sample preparation}

\begin{figure}[htbp]
\centering
  \includegraphics[width=0.45\columnwidth]{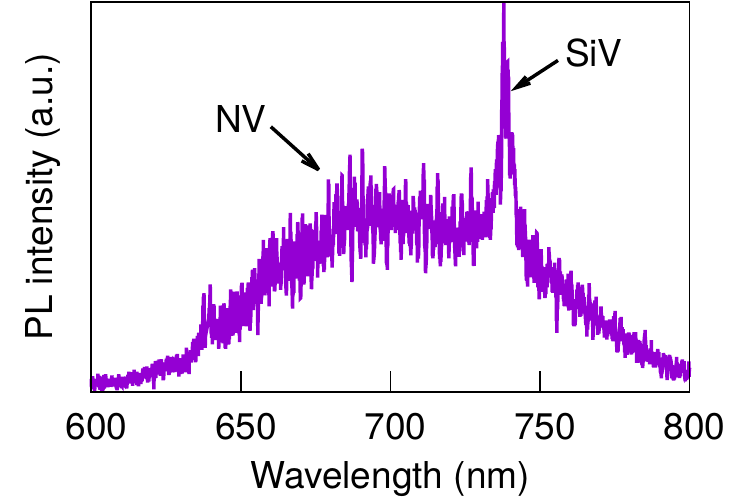}
  \caption{\label{fig:pl_3}
  Typical PL spectrum from samples~C, D and E, showing spectral signatures of both SiV centers (ZPL at 737\,nm) and nitrogen-vacancy (NV) centers (ZPL at 638\,nm, broad phonon sideband).
  }
\end{figure}

We study several ND samples grown using a HPHT process described in~\cite{davydov2014production} under varying conditions (see Table~\ref{tab:table2}).
The PL spectra of samples~C, D and E do not show a clean SiV signature, but instead contain contributions from both SiV and nitrogen-vacancy (NV) centers. A typical example is shown in Figure~\ref{fig:pl_3}.
It appears that the longer isothermal exposures for these samples resulted in a much higher abundance of NV centers compared to samples~A and B, 
and thus we are unable to obtain fluorescence that is emitted only from SiV centers.
Therefore, samples~C, D and E were not studied further in detail, and the remainder of this work focuses on samples~A and B.

Samples~A and B are prepared differently after extraction from the high-pressure apparatus.
Sample~A is boiled with a mixture of sulfuric, hydrochloric, and nitric aicds for 4\,h at 150\degree C to remove the graphite content.
After washing with distilled water, the sample is centrifuged (5000\,g 10\,min) in 100\% ethanol, the supernatant is extracted and centrifuged again.
The precipitate of the second centrifugation is suspended in isopropanol and ultrasonicated for 30\,mins.
10\,$\mu$l of the sample suspension is then spin-coated onto a silicon substrate.

Sample~B is not treated with acid boiling; it is suspended in isopropanol, ultrasonicated for 30 mins, then centrifuged (1900\,g 60\,min).
The top portion (3\,ml out of 12\,ml) is extracted, and
10\,$\mu$l of the sample suspension is spin-coated onto a silicon substrate.

\begin{table}[htbp]
\setlength{\tabcolsep}{10pt}
\caption{\label{tab:table2}
  Growth conditions for the different nanodiamond samples, detailing the Si/C ratio of the initial compound, pressure, and the temperature and duation of isothermal exposure.
  }
  \centering
  \begin{tabular}{clcl}
  Sample & Si/C ratio & Pressure & \quad Isothermal exposure\\
  \hline
  A & 0.008 & 8.0\,GPa & \quad  1300\degree C, 5\,s\\
  B & 0.05 & 8.0\,GPa &  \quad  1450\degree C, 5\,s\\
  C & 0.05 & 8.0\,GPa &  \quad  1300\degree C, 15\,s\\
  D & 0.08 & 8.0\,GPa &  \quad  1300\degree C, 15\,s\\
  E & 0.08 & 8.0\,GPa &  \quad  1300\degree C, 20\,s
\end{tabular}
\end{table}

\section{Appendix B: Experimental setup}

\begin{figure}[tbp]
\centering
  \includegraphics[width=0.65\columnwidth]{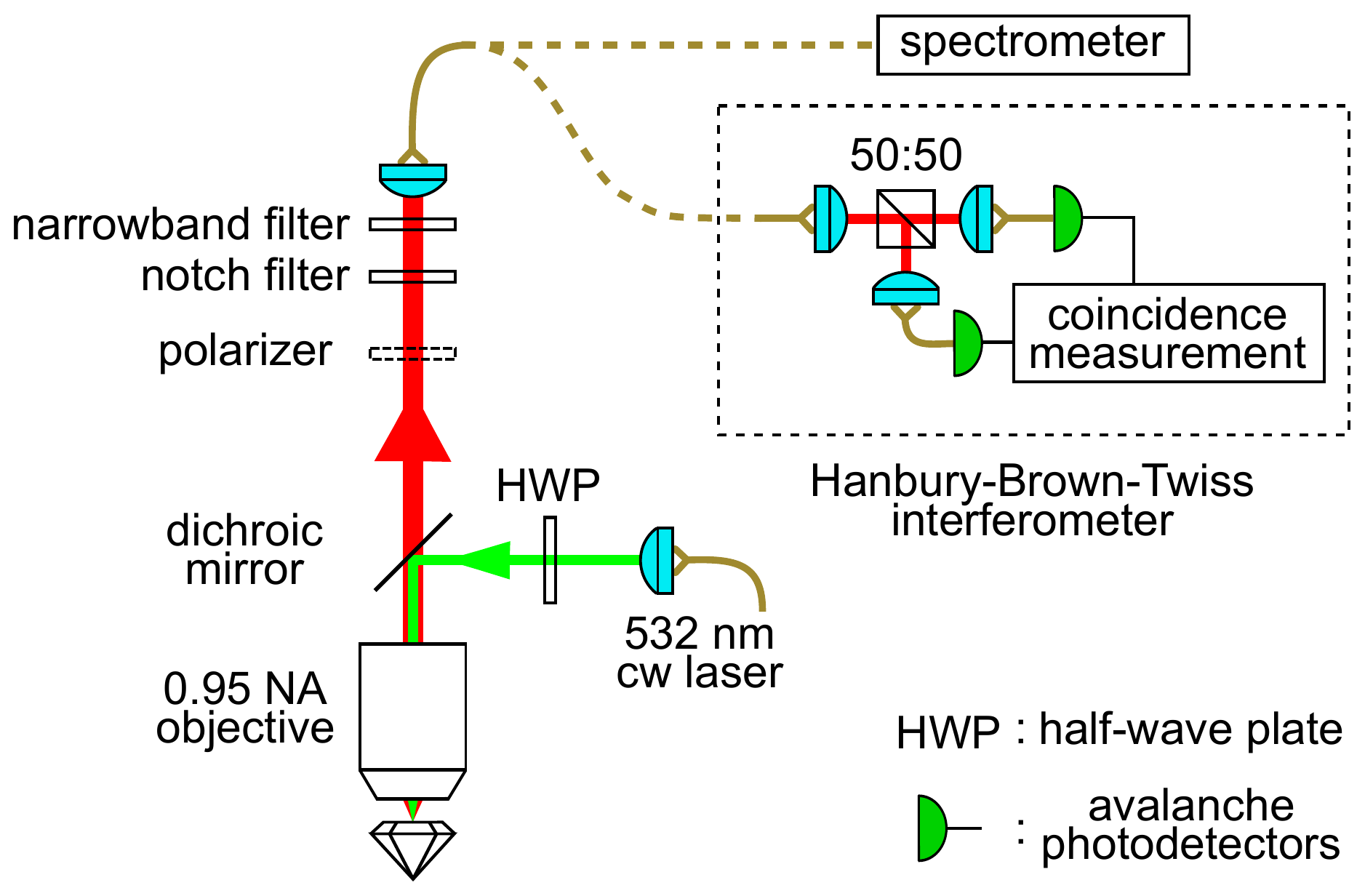}
  \caption{\label{fig:confocal}
  Sketch of the experimental setup. }
\end{figure}

Optical characterization is performed with a home-built room-temperature confocal microscope (see Fig.~\ref{fig:confocal}).
The sample is mounted on a 3D piezoelectric stage (Piezosystem Jena, TRITOR 100 SG), which is scanned in 0.5\,$\mu$m steps.
Off-resonant excitation light from a continuous-wave 532\,nm laser (Oxxius) is focused onto the sample through an air objective (Nikon, NA=0.95).
A 532\,nm half-wave plate in the excitation arm is rotated to best align the polarization of the excitation laser to the SiV dipole orientation, and thus maximize the amount of fluorescence.
The emission is filtered by a dichroic mirror (Semrock FF555-Di03), a notch filter (Semrock NF03-532E), and a narrowband filter (Semrock FF01-740/13); 
these filters reject the excitation light and transmit only in the vicinity of the ZPL.
We choose to collect the emission into a single-mode fiber (Thorlabs SM600) to minimize the collection of background fluorescence.

Photoluminescence (PL) measurements are performed with a grating spectrometer (Princeton Instruments IsoPlane 160, 0.07\,nm resolution) with the narrowband filter removed.
All other measurements are performed with the Hanbury-Brown and Twiss (HBT) interferometer,
which consists of a 50/50 non-polarizing beamsplitter with two avalanche photodetectors (APDs, Perkin Elmer SPCM-AQRH-15) at its outputs.
The reported count rates are a sum of the signals from both APDs; the APD signals are also timestamped (qutools quTAU) and analyzed to obtain the $g^{(2)}$ function.
For polarization measurements, a polarizer (Thorlabs LPVIS050-MP2) is placed in the emission arm and rotated while monitoring the APD count rates.

\section{Appendix C: Considerations of timing jitter}

The measured $g^{(2)}$ function is a convolution of the actual photon statistics of the ND fluorescence and the timing response of the setup.
As the timing jitter of each APD (350\,ps, from the datasheet) is already comparable to the excited state lifetime ($\sim$1\,ns), this would be a non-negligible contribution to the measurement results.

\begin{figure}[hbtp]
\centering
  \includegraphics[width=0.7\columnwidth]{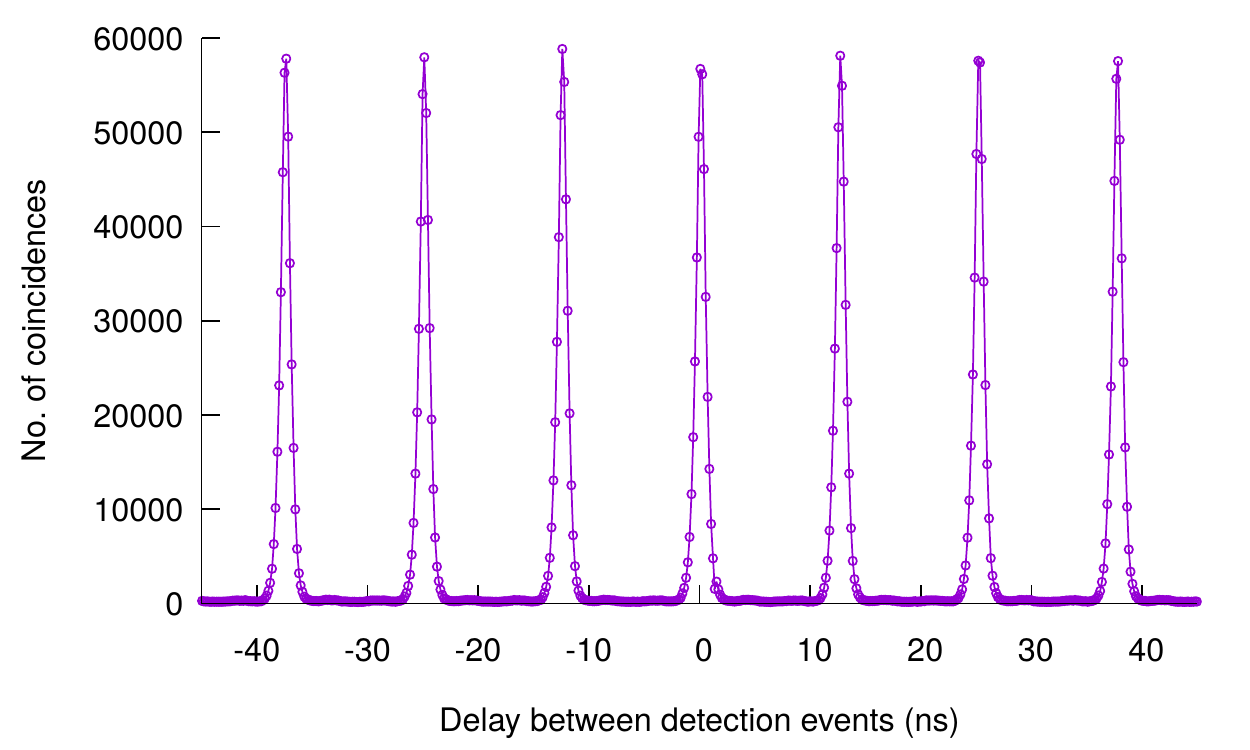}
  \caption{\label{fig:g2_suppl}
  Measured coincidences between the two APDs of the HBT setup, with 810\,nm femtosecond laser pulses as the input. The data is sorted into 162\,ps wide time bins.
  }
\end{figure}

We explicitly measure the timing response of the HBT setup by using attenuated 810\,nm femtosecond laser pulses as an input.
As the width of the laser pulses is negligible compared to the APD jitter, the width of the peaks in the coincidences between the APDs (see Fig.~\ref{fig:g2_suppl}) is a direct measure of the overall timing response of the setup.
The coincidence peaks are well-approximated by a Gaussian distribution function with a standard deviation $\sigma=493\pm1$\,ps, obtained by averaging the fits results of multiple peaks.

The full analytical expression used to fit the $g^{(2)}$ data is a convolution of 
$g_\textrm{noisy}^{(2)}(\tau)$ (defined in main text) with a Gaussian function, yielding:
\begin{equation}
  g_\textrm{fit}^{(2)}(\tau) = 1 - \frac{\rho^2}{2}\Big[(1+\alpha)f(\tau,\tau_1) - \alpha f(\tau,\tau_2) \Big]
\end{equation}
where
\begin{equation}
  f(\tau,\tau_{1,2}) = \exp(\frac{\sigma^2/\tau_{1,2} - 2\tau}{2\tau_{1,2}})\textrm{Erfc}(\frac{\sigma/\tau_{1,2} - \tau/\sigma}{\sqrt{2}})
  \;+\; \exp(\frac{\sigma^2+2\tau\tau_{1,2}}{2\tau_{1,2}^2})\textrm{Erfc}(\frac{\sigma/\tau_{1,2} + \tau/\sigma}{\sqrt{2}})
\end{equation}
and Erfc is the complementary error function.

\section*{Funding}
This work was supported by NRF-CRP14-2014-04, ``Engineering of a Scalable Photonics Platform for Quantum Enabled Technologies''.
VAD thanks the Russian Foundation for Basic Research (Grant No. 15-03-04490) for financial support.

\section*{Disclosures}
The authors declare that there are no conflicts of interest related to this article.
\end{document}